\documentclass{article}
\usepackage{graphicx}
\usepackage{epstopdf}
\usepackage{subfig}
\usepackage{amsfonts,amsmath}
\usepackage{cite}
\usepackage{hyperref}
\usepackage{setspace}
\usepackage{color}
\date{}

\begin{document}

\begin{tabular}{ p{\textwidth} }
\begin{center}
\begin{doublespace}
{\LARGE Noisy Classical Field Theories with Two Coupled Fields: Dependence of Escape Rates on Relative Field Stiffnesses }\\
{\large \bf Lan Gong$^{1}$ and D. L. Stein$^{1,2}$}\\
{\small \tt \href{mailto:lan.gong@nyu.edu}{lan.gong@nyu.edu}} ~ {\small \tt \href{mailto:daniel.stein@nyu.edu}{daniel.stein@nyu.edu}}\\
\end{doublespace}
$^{1}${\small \sl Department\ of Physics,~New York University,~New York,~NY 10003}\\
$^{2}${\small \sl Courant Institute of Mathematical Sciences,~New York University,~New York,~NY 10003}
\end{center}
\end{tabular}

\begin{abstract}
Exit times for stochastic Ginzburg-Landau classical field theories with two
or more coupled classical fields depend on the interval length on which the
fields are defined, the potential in which the fields deterministically
evolve, and the relative stiffness of the fields themselves. The latter is
of particular importance in that physical applications will generally
require different relative stiffnesses, but the effect of varying field
stiffnesses has not heretofore been studied. In this paper, we explore the
complete phase diagram of escape times as they depend on the various
problem parameters. In addition to finding a transition in escape rates as
the relative stiffness varies, we also observe a critical slowing down of
the string method algorithm as criticality is approached.
\end{abstract}
\small
\renewcommand{\baselinestretch}{1.25}
\normalsize

\section{Introduction}
\label{sec:introduction}

In a previous paper~\cite{GS10} (hereafter GS), the authors introduced and
solved a system of two coupled nonlinear stochastic partial differential
equations. Such equations are useful for modelling noise-induced activation
processes of spatially varying systems with multiple basins of attraction.
Examples of such processes include micromagnetic domain
reversal~\cite{MSK05,MSK06}, pattern
nucleation~\cite{Cross93,Tu97,Bisang98}, transitions in hydrogen-bonded
ferroelectrics~\cite{Dikande97}, dislocation motion across Peierls
barriers~\cite{GB97}, and structural transitions in monovalent metallic
nanowires~\cite{BSS05,BSS06}.  It is the last problem in particular that
the model introduced in GS was constructed to analyze.

The GS model provided a mathematical realization of a stochastic
Ginzburg-Landau field theory consisting of two coupled classical fields,
denoted $\phi_1(z)$ and $\phi_2(z)$, defined on a linear domain of finite
extent $L$. Stochastic partial differential equations of this type are
constructed to model noise-driven transitions between locally stable
states. In the especially important case of weak noise, where the
transition rate is of the Arrhenius form $\Gamma_0e^{-\Delta E/\epsilon}$,
with prefactor $\Gamma_0$ and activation barrier $\Delta E$ independent of
the noise $\epsilon$, the transition path occurs near (i.e., within a
lengthscale of order $O(\epsilon^{1/2})$) the saddle (or col) of least
action connecting the two stable states.

The two-field model displayed several interesting features, including a
type of ``phase transition'' in activation behavior as $L$ varied.  The
transition was driven by a change in the saddle state, from a uniform
configuration at small $L$ to a spatially varying one (``instanton'') at
larger $L$. This transition has been noticed and analyzed for the case of a
single field~\cite{MS01b,MS03}, but had not been seen in the rarely studied
case of a system with {\it two\/} coupled fields.  Perhaps more remarkably,
the system admitted an exact solution for the instanton state; such exact
solutions are rare in the case of nonlinear field theories with a single
field, much less a nontrivial system of coupled fields.

The introduction of two fields was required to study transitions among
different quantized conductance states in non-axisymmetric nanowires. The
axisymmetric case had previously been treated theoretically
in~\cite{BSS04,BSS05}. However, detailed studies using linear stability
analysis by Urban {\it et al.\/}~\cite{UBZSG04} indicated that roughly 1/4
of all such transitions involved either non-axisymmetric initial or final
states, or else a least-action transition passing through a
non-axisymmetric saddle. To describe such transitions, one field
($\phi_1(z)$) describes radial variations along the wire length and the
other ($\phi_2(z)$) describes deviations from axisymmetry.

One restriction of the analysis in GS was that the respective bending
coefficients $\kappa_1$ and $\kappa_2$ of the two fields were taken to be
equal. However, this is generally not the case in real
nanowires~\cite{UBZSG04}.  Therefore, in order to apply the model to actual
transitions, as well as to provide a complete picture of the activation
behavior in such systems, we need to consider the case where
$\kappa_1\ne\kappa_2$. In such cases analytical solutions cannot be found
and we need to rely on numerical methods. The study of this more general
problem is the subject of this paper.

\section{The Model}
\label{sec:model}
Consider two coupled classical fields $\phi_1(z)$, $\phi_2(z)$ on the interval $[-L/2,L/2]$, subject to the 
energy functional
\begin{align}
{\cal H}=\int_{-L/2}^{L/2} \!
(\frac{1}{2}\kappa_{1}(\phi_{1}'(z))^{2}+\frac{1}{2}\kappa_{2}(\phi_{2}'(z))^{2}+U(\phi_{1},\phi_{2}))
\, dz\, .
\label{eq:H}
\intertext{where}
U(\phi_1,\phi_2)=-\frac{\mu_1}{2}\phi_1^{2}+\frac{1}{4}\phi_1^{4}-\frac{\mu_2}{2}\phi_{2}^{2}+\frac{1}{4}\phi_2^{4}+\frac{1}{2}\phi_{1}^{2}\phi_{2}^{2}
\label{eq:U}
\end{align}

The bending coefficients $\kappa_1$, $\kappa_2$ can be related to the wire
surface tension~\cite{BSS04,BSS05}.  The arbitrary positive constants
$\mu_1$, $\mu_2$ are chosen such that $\mu_1 \neq \mu_2$, breaking
rotational symmetry. (The case $\mu_1=\mu_2$ has been investigated
analytically by Tarlie~{\it et al.}~\cite{Goldbart94} in the context of
phase slippage in conventional superconductors.)

If the system is subject to additive spatiotemporal white noise, 
then their time evolution is governed by the pair of stochastic partial differential equations:
\begin{eqnarray}
\label{eq:timeevolution}
\dot{\phi_{1}}=\kappa_1\phi_{1}''+\mu_{1}\phi_{1}-\phi_{1}^{3}-\phi_{1}\phi_{2}^{2}+\sqrt{2\epsilon}\,\xi_{1}(z,t)\nonumber\\
\dot{\phi_{2}}=\kappa_2\phi_{2}''+\mu_{2}\phi_{2}-\phi_{2}^{3}-\phi_{1}^{2}\phi_{2}+\sqrt{2\epsilon}\,\xi_{2}(z,t)\, ,
\end{eqnarray} 
where $\xi_{1,2}$ are the spatiotemporal noise terms satisfying
$<\xi_{i}(z_{1},t_{1})\xi_{j}(z_{2},t_{2})>=\delta(z_{1}-z_{2})\delta(t_{1}-t_{2})\delta_{ij},~i,j=1,2$.
If the noise is due to thermal fluctuations,
then by the fluctuation-dissipation theorem $\epsilon = k_BT$.

The activation energy $\Delta E$ and prefactor $\Gamma_0$ in the Arrhenius
rate formula depend not only on the details of the potential~(\ref{eq:U}),
but also on the interval length~$L$ on which the fields are defined, and on
the choice of boundary conditions at the endpoints $z=-L/2$ and~$z=L/2$.
It was shown in~\cite{Burki03} that Neumann boundary conditions are
appropriate for the nanowire problem, and we will employ them here.

In the usual case of a single field, the bending coefficient $\kappa$
plays a role in determining the intrinsic lengthscale (and therefore the
transition length at which the saddle state changes) on which field
variations occur; but once it is absorbed into a dimensionless lengthscale
by rescaling along with the parameters determining the potentials, it plays
no further role. Now, however, there are {\it two\/} bending coefficients,
and varying their relative magnitudes can in principle lead to new
phenomena. The aim of this paper is to study the effects of these
variations.

The metastable and saddle states are time-independent solutions of the zero-noise equations:
\begin{eqnarray}
\kappa_1\phi_1''=-\mu_1\phi_1+\phi_1^{3}+\phi_1\phi_2^{2}\nonumber\\
\kappa_2\phi_2''=-\mu_2\phi_2+\phi_2^{3}+\phi_2\phi_1^{2}
\label{eq:zero-noise}
\end{eqnarray}
Without loss of generality, we choose $\mu_{1}>\mu_{2}$.

Then there are two metastable states: $\phi_{1,s}=\pm\sqrt{\mu_1}$,
$\phi_{2,s}=0$; two spatially uniform saddle states: $\phi_{1,u}=0$,
$\phi_{2,u}=\pm\sqrt{\mu_2}$; and spatially nonuniform saddle states, or
instantons. When $\kappa_1=\kappa_2(=1)$, analytical solutions for the
instanton saddle states can be found:
\begin{gather}
\phi^{\rm
inst}_{1,m}(z)=\pm\sqrt{m}\sqrt{(2\mu_{1}-\mu_{2})-m(\mu_{1}-\mu_{2})}{\rm sn}(\sqrt{\mu_{1}-\mu_{2}}\,z|m)
\label{eq:ins1}\\ \phi^{\rm
inst}_{2,m}(z)=\pm\sqrt{\mu_{2}-m(\mu_{1}-\mu_{2})}{\rm dn}(\sqrt{\mu_{1}-\mu_{2}}\,z|m)
\label{eq:ins2}
\end{gather}
where ${\rm sn}(.|m)$and ${\rm dn}(.|m)$ are the Jacobi elliptic
functions with parameter $m$, whose periods are $4K(m)$ and $2K(m)$
respectively, with $K(m)$ the complete elliptic integral of the first
kind~\cite{Abramowitz65}. The parameter $m \in [0, 1]$ is related to
interval length $L$ through the Neumann boundary conditions, with
$m\to 0^+$ corresponding to $L\to L_c^+$, where $L_c$ is the critical
length, and $m\to 1$ corresponding to
$L\to\infty$~\cite{MS01b,MS03,GS10}. When $\kappa_1=\kappa_2=1$,
\begin{equation} 
\label{eq:lc}
L=\frac{2K(m)}{\sqrt{\mu_{1}-\mu_{2}}}
\end{equation}

We found in GS that varying $L$ triggers a transition between the
uniform and instanton saddle states; the critical length $L_c$ is
determined by:
\begin{equation}
\label{eq:Lc}
L_c = \frac{2K(0)}{\sqrt{\mu_{1}-\mu_{2}}}\\
=\frac{\pi}{\sqrt{\mu_{1}-\mu_2}}
\end{equation}
This results in a transition in the activation
behavior, including anomalous behavior at the critical length.  Such a
transition may have already been seen experimentally~\cite{BSS06}, in a
crossover from ohmic to nonohmic conductance as the voltage across gold
quantum point contacts increases~\cite{YOT05}.  We will show below that the
same effect occurs when the ratio~$\kappa_1/\kappa_2$ is varied.

As noted above, the transition rate in the low-noise ($\epsilon\to 0$)
limit is given by the Kramers formula:
\begin{equation}
\label{eq:rate}
\Gamma\sim\Gamma_0 \exp(-\Delta E /\epsilon)\, .
\end{equation}
Here $\Delta E$ is the activation barrier, that is, the difference in
energy between the saddle and the starting metastable states, while
$\Gamma_0$ is the rate prefactor:
\begin{equation}
\label{eq:prefactor}
\Gamma_{0}=\frac{1}{\pi}\sqrt{\left|\frac{\det\,{\bf\Lambda}_{s}}{\det\,{\bf\Lambda}_{u}}\right|}|\lambda_{u,1}|\, .
\end{equation}
In the above equation ${\bf\Lambda}_s$ is the linearized dynamical operator
describing perturbations about the metastable state; similarly
${\bf\Lambda}_u$ describes perturbations about the saddle.  $\lambda_{u,1}$
is the single negative eigenvalue of ${\bf\Lambda}_{u}$, corresponding to
the direction along which the most probable transition path approaches the
saddle state.  The behavior of $\Gamma_0$ becomes anomalous at the critical
point $L_c$, where fluctuations around the most probable path become large.

\section{Calculation of the Minimum Energy Path}
\label{sec:mep}

Computation of exit behavior requires knowledge of the transition path(s),
in particular behavior near the local minimum and the saddle. In our model,
both are found as solutions of two coupled nonlinear differential
equations~\cite{GS10}. A powerful numerical technique constructed
explicitly for this type of problem is the ``string method'' of E, Ren, and
Vanden-Eijnden\cite{ERenEric02,ERenEric07}.  The algorithm proceeds by
evolving smooth curves, or strings, under the zero-noise dynamics.  These
strings connect the beginning and final locally stable states, and in
between the two ends each string contains a series of intermediate states
called "images". The method is constructed so that the string evolves
towards the most probable transition path.  The evolution proceeds until
the condition for equilibrium is reached:
\begin{equation}
[\delta{\cal H}]^{\perp}=0
\end{equation}
\label{eq:string equation}
where ${\cal H}$ is given by~(\ref{eq:H}) and $[\delta{\cal H}]^{\perp}$ is its component 
perpendicular to the string. 

Once equilibrium is reached, the string images correspond to the
configurations sampled by the system at different stages of the activation
process.  The image with highest energy is the one nearest the saddle
state. In order to get an accurate result, the distribution of images needs
to be sufficiently fine-grained.  In our computation, we used 61 images
(including the two ends of the string); because of the symmetry of our
energy functional, the image in the middle corresponds to the saddle.

When such symmetry is absent and the location of the saddle needs to be
determined with high precision, one can use an alternative method to the
brute force one of simply increasing the number of images along the
string. This alternative requires first finding a rough approximation of
the most probable transition path, again using the string method but with a
small number of images, and then switching to a "climbing image" algorithm
in which one picks up an image that is believed close to the saddle and
then drive it towards the saddle. The climbing force is obtained from
inverting the energy gradient along the direction of the unstable
eigenvector of the saddle state.  Details can be found
in~\cite{ERenEric02,ERenEric07}.

We have found an analogue to critical slowing down in the current context:
near criticality convergence of the string method becomes increasingly
slow.  Expanding the energy functional around the saddle reveals that the
lowest stable eigenvalue vanishes to second order, leading to a rapid
increase in relaxation time. This phenomenon will be further investigated
in the following sections.

\section{Results}
\label{sec:results}

We now turn to the case $\kappa_{1} \neq \kappa_{2}$. To begin, we fix
$\kappa_{2}=1$ and vary $\kappa_{1}$.  We consider the cases where
$\kappa_1$ is both less than and greater than 1. Because the critical
length now depends on $\kappa_1$, we denote it $L_c(\kappa_1)$.

\begin{figure}[!htp]
\centering
\subfloat[$\phi^{\rm saddle}_{1}(z)$]{\label{fig:trans1 evolution}\includegraphics[width=0.45\textwidth]{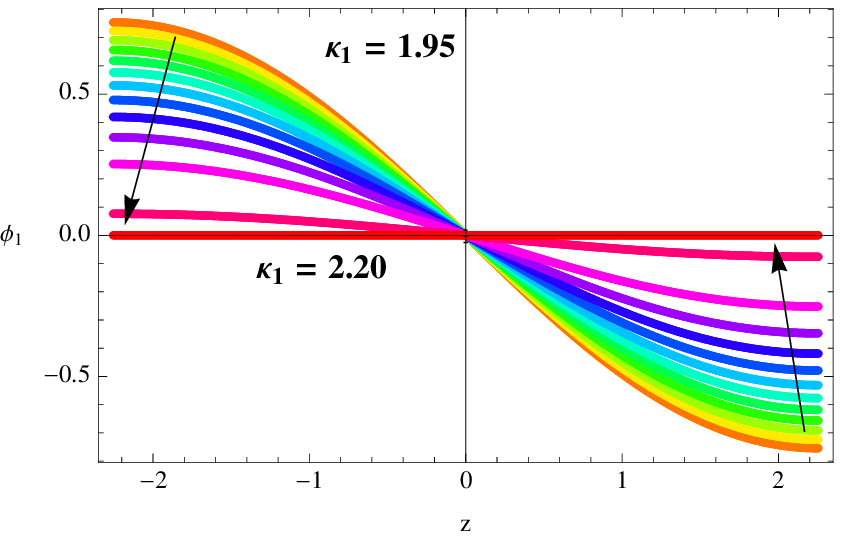}}\quad
\subfloat[$\phi^{\rm saddle}_{2}(z)$]{\label{fig:trans2 evolution}\includegraphics[width=0.45\textwidth]{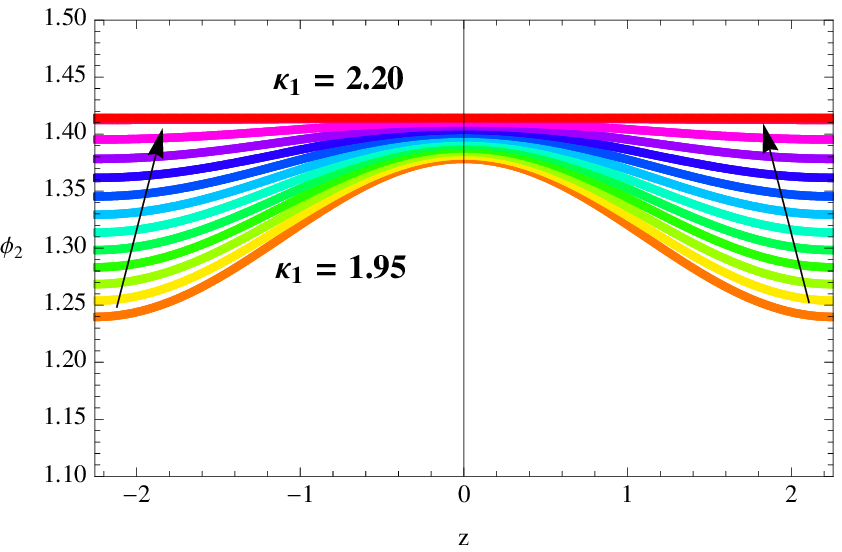}}
\caption{ The saddle states $\phi^{\rm saddle}_{i=1,2}$ passed by the system at different relative field stiffnesses ($=\kappa_1$). The evolution of the saddle is described as the continuous transition of colors. $\kappa_1$ was increased from 1.95 to 2.20 with an increment of 0.01. The arrows indicate the suppression of the instanton at $L>L_c(1)=\frac{\pi}{\sqrt{\mu1-\mu2}}$ as $\kappa_1$ increases. Here $\mu_{1}=3$, $\mu_{2}=2$, and $L=4.51$.}
\label{fig:transition state evolution}
\end{figure}

\begin{figure}[!htp]
\centering
\subfloat[$\phi^{\rm saddle}_{1}(z)$]{\label{fig:trans1 evolution 250}\includegraphics[width=0.45\textwidth]{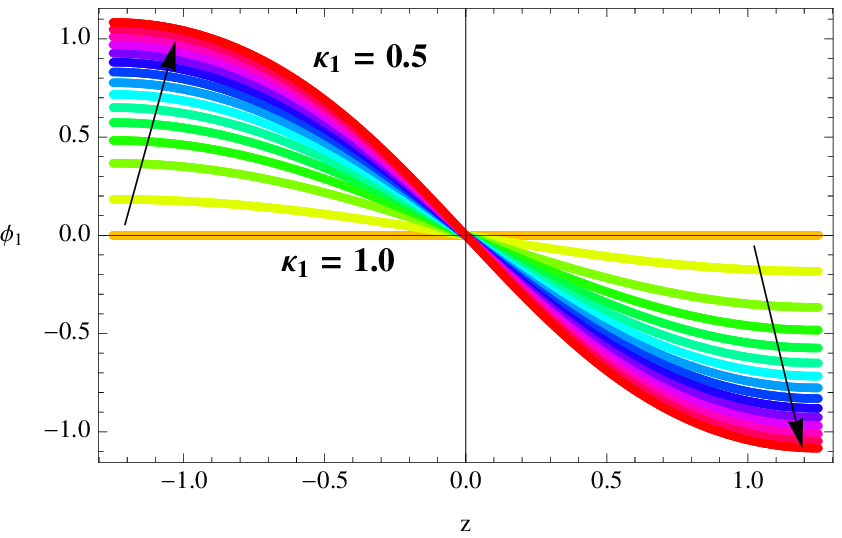}}\quad
\subfloat[$\phi^{\rm saddle}_{2}(z)$]{\label{fig:trans2 evolution 250}\includegraphics[width=0.45\textwidth]{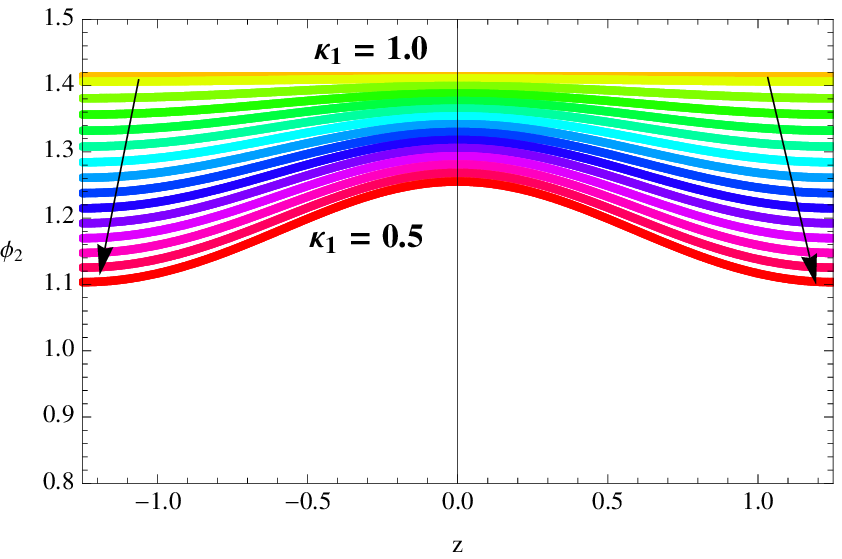}}
\caption{The saddle states $\phi^{\rm saddle}_{i=1,2}$ passed by the system at different relative field stiffnesses(=$\kappa_1$). The evolution of the saddle is described as the continuous transition of colors. $\kappa_1$ was decreased from 1.0 to 0.5 with a decrement of -0.01. The arrows indicate the retrieval of the instanton at $L<L_c(1)=\frac{\pi}{\sqrt{\mu1-\mu2}}$ as $\kappa_1$ decreases. Here $\mu_{1}=3$, $\mu_{2}=2$, and $L=0.25$.}
\label{fig:transition state evolution 250}
\end{figure}

As noted earlier (cf.~(\ref{eq:Lc}),
$L_{c}(1)=\frac{\pi}{\sqrt{\mu_{1}-\mu_{2}}}$: below $L_c(1)$ the
saddle is spatially uniform, and above $L_c(1)$ it is spatially
varying~\cite{GS10}.  The situation becomes more complicated when
$\kappa_1 \neq 1$. Fig.~\ref{fig:transition state evolution}
summarizes our results when $\mu_1=3$, $\mu_2=2$ and $L>L_c(1)=\pi$.
In this and Fig.~\ref{fig:transition state evolution 250}, the saddle
state (whether uniform or instanton) is denoted $\phi_i^{\rm
  saddle}(z), i=1,2$.  We find that as $\kappa_{1}$ increases, the
spatial variation of the instanton becomes increasingly suppressed
until the instanton finally collapses to the uniform
state. Conversely, when $L<L_c(1)$, the instanton state is retrieved
for $\kappa_1<1$ (cf.~Fig.~\ref{fig:transition state evolution 250}).

\begin{figure}[!htp]
\centering
\subfloat[$L=2.50$]{\label{fig:energy250}\includegraphics[width=0.45\textwidth]{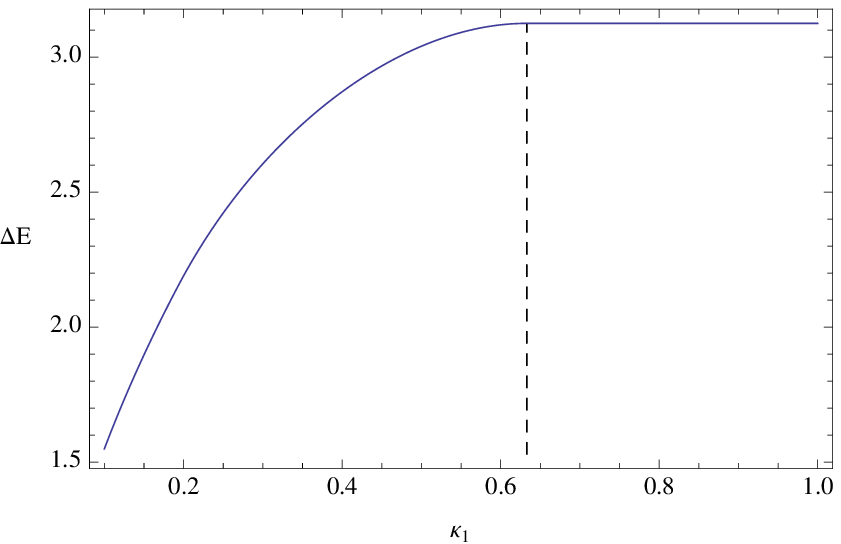}}\quad
\subfloat[$L=4.51$]{\label{fig:energy451}\includegraphics[width=0.45\textwidth]{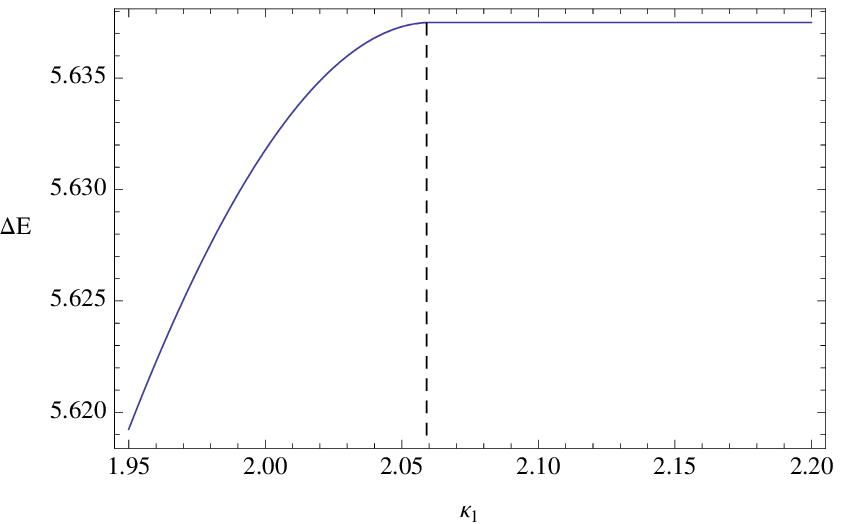}}
\caption{Energy of the saddle state as a function of $\kappa_{1}$,
  where $\Delta E=E_{saddle}-E_{metastable}$. The dashed lines
  indicate that $\kappa_{1c} \approx 0.63$ for $L=2.50$ and
  $\kappa_{1c} \approx 2.06$ for $L=4.51$, where energies of the
  instanton and uniform saddles cross. The region
  $\kappa_1<\kappa_{1c}(L)$ corresponds to the instanton saddle, and
  $\kappa_1>\kappa_{1c}(L)$ to the uniform saddle. Here $\mu_{1}=3$,
  and $\mu_{2}=2$.}
\label{fig:energy}
\end{figure}

This effect can be understood as follows. In the limit of low noise, the
transition occurs over the saddle state of least energy. An increase in
$\kappa_1$ raises the bending energy of any nonuniform configuration, and
therefore that of the instanton, while leaving the energy of the uniform
saddle unchanged. When the energies of these two states cross, the
activation process switches from one saddle state to the other.  This is
seen explicitly in Fig.~\ref{fig:energy}, where we plot the energy of the
saddle state against $\kappa_1$ for both $L=0.25$ and $L=4.51$.  In these
figures the curve to the left of the dashed line is the instanton branch,
which increases monotonically until it reaches a constant value: the energy
of the uniform saddle state.

We next investigated the question of whether the transition from uniform
saddle to instanton (or vice-versa) as $\kappa_1$ varies occurs as a
continuous crossover or as an abrupt phase transition.  If the latter, then
we also need to determine the order of the transition.

In~\cite{GS10}, the uniform~$\to$~instanton saddle transition was
induced by changing $L$ at fixed $\kappa_1$. There we concluded that
the transition was reminiscent of a second-order phase transition, in
the asymptotic $\epsilon\to 0$ limit. This follows from the continuity
of the activation energy at all values of $L$, including
$L_c(1)$. (For examples of potentials where the activation energy
jumps at a precise value of $L$, corresponding to a first-order
transition, see~\cite{BSS08}.)  In fact, it can be shown analytically
that the first derivative of the activation energy curve with respect
to $L$ is also continuous everywhere, but the second derivative is
discontinuous at $L_c(1)$.

\begin{figure}[!htp]
\centering
\includegraphics[width=\linewidth]{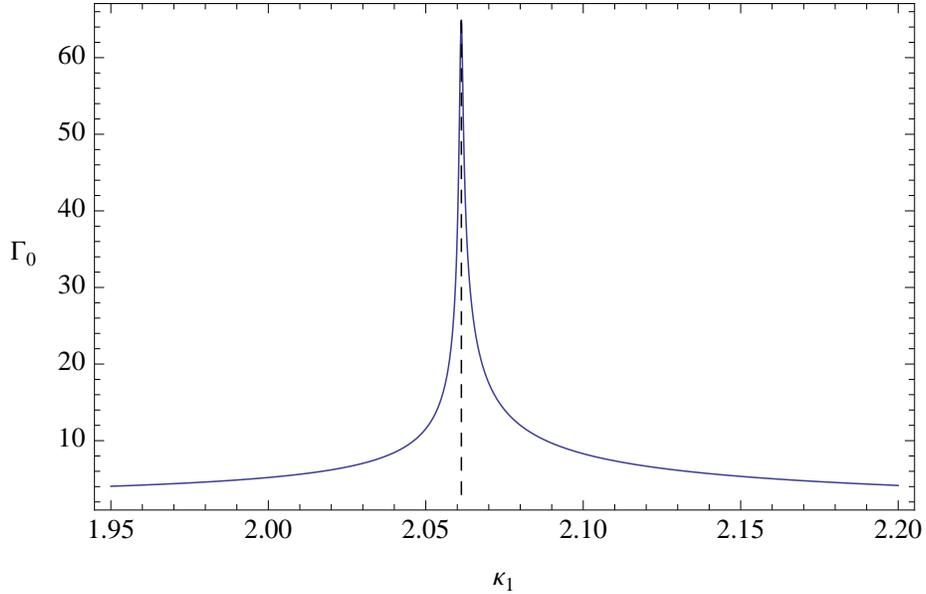}
\caption{Behavior of the prefactor $\Gamma_0$ as calculated numerically
  using~(\ref{eq:prefactor}) when $\mu_{1}=3$, $\mu_{2}=2$, and
    $L=4.51$.}
\label{fig:prefactor_L451}
\end{figure}

Similarly, Fig.~\ref{fig:energy} suggests that there is indeed a continuous
phase transition, in that the activation energy changes continuously as one
passes through the transition, as $\kappa_1$ varies for fixed $L$.  This
continuity leads to a divergence in the transition rate prefactor, shown in
Fig.~\ref{fig:prefactor_L451} (similar to that induced by changing $L$ at
fixed $\kappa_1$ in~\cite{GS10}).  The value of $\kappa_{1c}(L=4.51)$
where the prefactor diverges and that where the energies of the respective
saddles cross agree to within a numerical error of $10^{-2}$.

What causes this divergence? Away from criticality, the spectrum of
the linearized dynamical operator $\Lambda_u$ about the saddle
consists of a single negative eigenvalue, whose corresponding
eigenvector determines the unstable direction, with all other
eigenvalues positive. As criticality is approached, the smallest
positive eigenvalue, denoted $\lambda_{u,2}$, approaches zero. This
signals the mathematical divergence on the ``normal'' lengthscale of
$O(\epsilon^{1/2})$ of fluctuations about the saddle, and
by~(\ref{eq:prefactor}) is seen to lead to divergence of the
prefactor. (For a discussion of how to interpret this ``divergence'',
see~\cite{Stein04}.)

\begin{figure}[!htp]
\centering
\includegraphics[width=\linewidth]{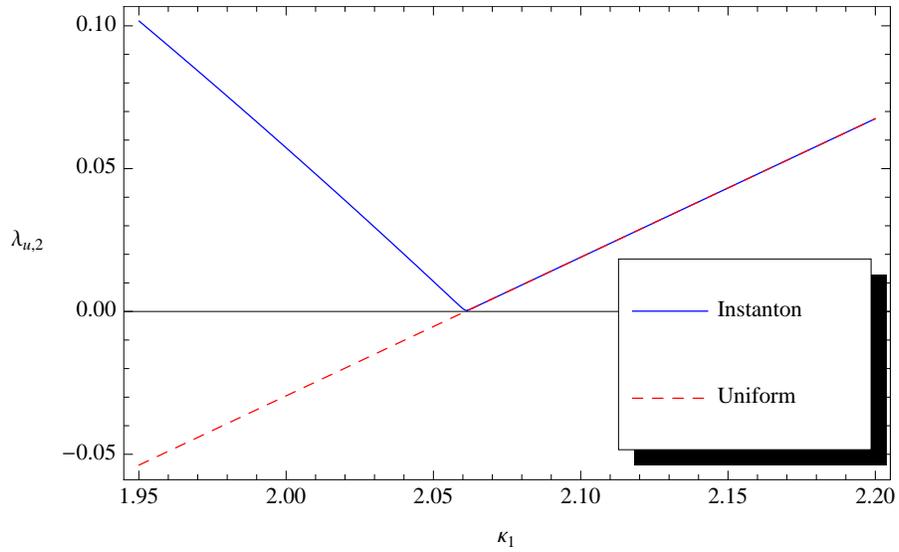}
\caption{Smallest nonnegative eigenvalue $\lambda_{u,2}$ (solid line)
  of the saddle state for $\mu_{1}=3$, $\mu_{2}=2$ and $L=4.51$. For
  $\kappa_1<2.06$ this eigenvalue corresponds to the instanton saddle,
  and for $\kappa_1>2.06$ to the uniform saddle. The extended dashed
  line shows the continuation of this eigenvalue for the uniform
  saddle in its unstable regime below $\kappa_{1c}$.}

\label{fig:lambda2}
\end{figure}

The eigenvalue spectrum about the uniform saddle can be analytically
calculated~\cite{GS10}. The eigenvalue $\lambda_{u,2}$ is found to be
\begin{equation}
\label{eq:lambda2}
\lambda_{u,2}=\frac{\kappa_1\pi^{2}}{L^{2}}-(\mu_1-\mu_2)\, .
\end{equation}
At fixed $L$, this switches from negative (unstable) to positive
(stable) as $\kappa_1$ increases, as shown in
Fig.~\ref{fig:lambda2}. This change of sign corresponds to a
transition from an instanton saddle to a uniform one as $\kappa_1$
varies. Using this approach, the curve $L_c$~vs.~$\kappa_{1c}$ can be
derived analytically as the locus of points where $\lambda_{u,2}=0$ 
and thus the full phase diagram determined as
represented by the solid curve in Fig.~\ref{fig:phasediagram}.

We have also studied the behavior of the transition rate prefactor in a
wide range of values of $L$ numerically, all of which lead to the same
conclusion as described above. Fig.~\ref{fig:prefactor_all} shows the
divergence of $\Gamma_0$ at different $L_c$'s and their corresponding
$\kappa_{1c}$'s.

\section{Discussion}
\label{sec:discussion}

We have solved the general two-field model of~(\ref{eq:H}) and~(\ref{eq:U})
for its full parameter space. We have uncovered a new mechanism for the
transition in the switching rate, and shown that it has features of a
second-order phase transition.

In the one-field case, the mechanism behind the transition is not difficult
to understand.  At smaller $L$ (recall that this is a {\it dimensionless}
lengthscale, in units of a reduced length that includes the single bending
coefficient $\kappa$), bending costs (in conformity with the boundary
condition) are prohibitive, and the uniform saddle therefore has lower
energy than the instanton. At large lengthscales, the uniform saddle has a
prohibitive bulk energy (i.e., potential difference with the stable state),
whereas the instanton differs from one or the other stable state only
within the domain wall region, whose lengthscale remains of $O(1)$. What is
perhaps less intuitive is that the transition in saddle states should be
asymptotically (as $\epsilon\to 0$) sharp.

\begin{figure}[!htp]
\centering
\includegraphics[width=\linewidth]{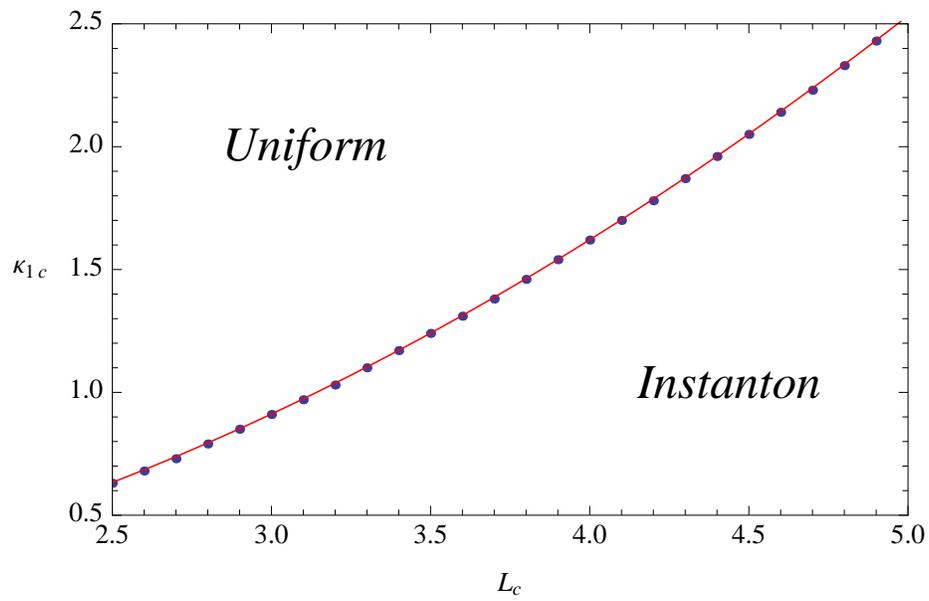}
\caption{The phase diagram at $\mu_{1}=3$, $\mu_{2}=2$. The dots
  represent numerically determined values where the energies of the
  uniform and instanton saddles cross.  The solid line was computed
  analytically using~(\ref{eq:lambda2}), that is, by determining the
  relation between $\kappa_1$ and $L$ along which the smallest
  nonnegative eigenvalue of the uniform saddle is zero.}
\label{fig:phasediagram}
\end{figure}

\begin{figure}[!htp]
\centering
\includegraphics[width=\linewidth]{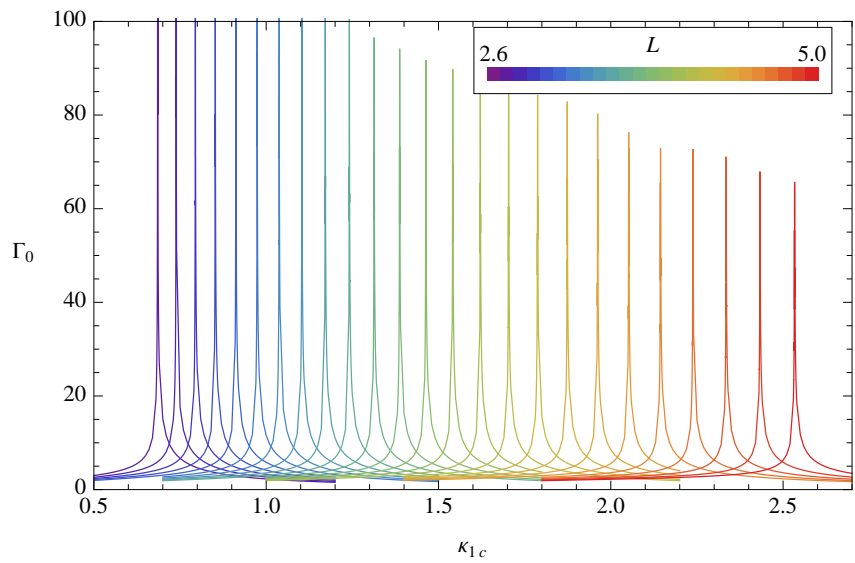}
\caption{The divergence of $\Gamma_0$ at $L$ ranging from $2.6$ to
  $5.0$. Here $\mu_{1}=3$, $\mu_{2}=2$, and the values of
  $\kappa_{1c}$ (or the corresponding $L_c$) correspond to the dots in
  Fig.~\ref{fig:phasediagram}.  The peaks are of different heights
  because the speeds of divergence are not necessarily the same for
  every $L$. }
\label{fig:prefactor_all}
\end{figure}

Here we have uncovered a second mechanism for the transition to occur: as
noted in Sec.~\ref{sec:results}, increase of $\kappa_1$ when $L>L_c(1)$
suppresses spatial variation, causing the instanton (again in a sharp
transition) to collapse to the uniform saddle.  Conversely, the instanton
state can be retrieved for $L<L_c(1)$ when $\kappa_1$ {\it decreases\/}; of
course, bending becomes increasingly favorable energetically. The result is
a phase diagram in $(L,\kappa_1/\kappa_2)$~space, as
in~Fig.~\ref{fig:phasediagram}, where a phase separation line denotes the
boundary between the uniform and instanton ``phases''.

We close with some remarks about the string method as applied to this problem.

A randomly placed string will relax towards the most probable transition
path along the stable direction of the saddle. In Sec.~\ref{sec:results} we
defined the smallest positive eigenvalue (corresponding to the stable
direction) of the linearized operator $\Lambda_u$.  As a second-order phase
transition is approached, $\lambda_{u,2}$ drops to $0^{+}$, so that the
energy landscape curvature in the stable direction becomes very small. When
the string arrives in its neighborhood, the restoring force exerted along
its normal direction correspondingly becomes small leading to slow
convergence.  If one sits right at the critical point, the string will not
arrive at the saddle.

The string method assumes that most of the probability flux from the
reactant to the product state is carried by one (or more generally a few)
paths through the saddle state, in each of which the probability flux is
tightly confined to a narrow quasi - 1D region in state space.  However,
near criticality the path splays out in the direction normal to the
longitudinal transition path.  In this case one needs to consider
transition ``tubes'', inside which most of probability flux is
concentrated. The equilibrium condition~(\ref{eq:string equation})
corresponds to conditions away from criticality, where the transition tube
is thin.

The equation for the path of maximum flux is derived in~\cite{EEric10},
where it is noted that the reaction flux intensity must be maximized along
the thin transition tube (or the string, when using the string method). An
alternative derivation can be found in~\cite{BerkowitzMorgan83}.

\smallskip

{\it Acknowledgments.\/} The authors are grateful to Weiqing Ren and Ning
Xuan for helpful discussions.  We are especially grateful to Gabriel Chaves
for his help in programming the string method.  This work was supported in
part by NSF Grant PHY-0965015.

\bibliographystyle{ieeetr}
\bibliography{Reference}

\clearpage

\end{document}